\documentclass[twocolumn]{aastex62}

\graphicspath{{./}{figures/}}

\received{October, 2019}
\accepted{January, 2020 }

\shorttitle{The nature of the puzzling SMG AzTEC2}
\shortauthors{Jim\'enez-Andrade et al.}

\begin{document}

\title{The redshift and  star formation mode of AzTEC2:  a pair of massive galaxies at $z=4.63$}

\correspondingauthor{E.~F. Jim\'enez-Andrade}
\email{ejimenez@nrao.edu}

\author[0000-0002-2640-5917]{E.~F. Jim\'enez-Andrade}
\affiliation{National Radio Astronomy Observatory, 520 Edgemont Road, Charlottesville, VA 22903, USA} 
\affil{Argelander Institute for Astronomy, University of Bonn, 
 Auf dem H\"ugel 71, D-53121 Bonn, Germany}

\author[0000-0002-7051-1100]{J.~A. Zavala}
\affiliation{Department of Astronomy, The University of Texas at Austin, 2515 Speedway Blvd Stop C1400, Austin, TX 78712, USA}

\author[0000-0002-6777-6490]{B. Magnelli}
\affil{Argelander Institute for Astronomy, University of Bonn, 
 Auf dem H\"ugel 71, D-53121 Bonn, Germany}
 
 \author[0000-0002-0930-6466]{C.~M. Casey}
\affiliation{Department of Astronomy, The University of Texas at Austin, 2515 Speedway Blvd Stop C1400, Austin, TX 78712, USA}

 \author[0000-0001-9773-7479]{D. Liu}
\affiliation{Max Planck Institut f\"ur Astronomie, K\"onigstuhl 17, D-69117, Heidelberg, Germany}

\author[0000-0002-0071-3217]{E. Romano-D\'iaz}
\affil{Argelander Institute for Astronomy, University of Bonn, 
 Auf dem H\"ugel 71, D-53121 Bonn, Germany}

 \author[0000-0002-3933-7677]{E. Schinnerer}
\affiliation{Max Planck Institut f\"ur Astronomie, K\"onigstuhl 17, D-69117, Heidelberg, Germany}

 \author[0000-0001-5429-5762]{K. Harrington}
\affil{Argelander Institute for Astronomy, University of Bonn, 
 Auf dem H\"ugel 71, D-53121 Bonn, Germany}

 \author[0000-0002-6590-3994]{I. Aretxaga}
\affiliation{Instituto Nacional de Astrof\'isica, \'Optica y Electr\'onica (INAOE), Luis Enrique Erro 1, Sta. Ma. Tonantzintla, Puebla, M\'exico}

\author[0000-0002-8414-9579]{A. Karim}
\affil{Argelander Institute for Astronomy, University of Bonn, 
 Auf dem H\"ugel 71, D-53121 Bonn, Germany}

\author[0000-0002-8437-0433]{J. Staguhn}
\affil{The Henry A. Rowland Department of Physics and Astronomy, Johns Hopkins University, 3400 North Charles Street, Baltimore, MD 21218, USA}
\affil{Observational Cosmology Lab, Code 665, NASA Goddard Space Flight Center, Greenbelt, MD 20771, USA}

\author{A.~D. Burnham}
\affiliation{Department of Astronomy, The University of Texas at Austin, 2515 Speedway Blvd Stop C1400, Austin, TX 78712, USA}

 \author[0000-0003-4229-381X]{A. Monta\~na}
\affiliation{CONACyT - Instituto Nacional de Astrof\'isica, \'Optica y Electr\'onica, Luis E. Erro 1, Tonantzintla, Puebla, M\'exico}

\author{V. Smol\v{c}i\'c}
\affiliation{Faculty of Science, University of Zagreb, Bijeni\v{c}ka c. 32, 10002 Zagreb, Croatia}

\author[0000-0001-7095-7543]{M. Yun}
\affil{Department of Astronomy, University of Massachusetts, Amherst, MA 01003, USA}

\author[0000-0002-1707-1775]{F. Bertoldi}
\affil{Argelander Institute for Astronomy, University of Bonn, 
 Auf dem H\"ugel 71, D-53121 Bonn, Germany}

  \author{D. Hughes}
\affiliation{Instituto Nacional de Astrof\'isica, \'Optica y Electr\'onica (INAOE), Luis Enrique Erro 1, Sta. Ma. Tonantzintla, Puebla, M\'exico}



\begin{abstract}

We combine observations from the Atacama Large Millimeter/submillimeter Array (ALMA) and the  NOrthern Extended Millimeter Array (NOEMA) to  assess the redshift and to study the star formation conditions in AzTEC2: one of the brightest sub-millimeter  galaxies (SMGs) in the COSMOS field ($S_{\rm 1.1mm}=10.5\pm1.4$\,mJy). Our high-resolution observations confirm that AzTEC2 splits into two components (namely AzTEC2-A and AzTEC2-B) for which we detect [C\,II] and $^{12}$CO(5$\to$4) line emission, implying a redshift of $4.626\pm0.001$ ($4.633\pm0.001$) for AzTEC2-A (AzTEC2-B) and ruling out previous associations with a galaxy at $z\sim1$.
We use the $^{12}$CO(5$\to$4) line emission and adopt typical SMG-like gas excitation conditions to  estimate the molecular gas mass, which is $M_{\rm gas}(\alpha_{\rm  CO}/2.5)=2.1\pm0.4 \times10^{11}\,M_\odot$\, for  AzTEC2-A, and a factor four lower for AzTEC2-B. With the infrared-derived star formation rate of  AzTEC2-A ($1920\pm100 \,M_\odot$\,yr$^{-1}$) and AzTEC2-B ($710\pm 35\,M_\odot$\,yr$^{-1}$), they both will consume their current gas reservoir within $(30-200)$\,Myr.  We find  evidence of a rotation-dominated [C\,II] disk in AzTEC2-A, with a de-projected rotational velocity of  $v_{\rm rot}(i=39^\circ)=660\pm130$\,km\,s$^{-1}$, velocity dispersion $\lesssim100$\,km\,s$^{-1}$, and dynamical mass of $M_{\rm dyn}(i=39^\circ)=2.6^{+1.2}_{-0.9}\times10^{11}\,M_\odot$.  We  propose that an elevated gas accretion rate  from the cosmic web might be the main driver of the intense levels of star formation in AzTEC2-A,  which might be further enhanced by gravitational torques induced by its minor companion (AzTEC2-B). These results strengthen the picture whereby  the population of single-dish selected SMGs  is rather heterogeneous,  including a   population of  pairs of massive, highly-active galaxies in a pre-coalescence phase.

\end{abstract}



\section{Introduction} \label{sec:intro}

Empirical and theoretical evidence indicate that the global production of stars in galaxies is mainly regulated by the  steady accretion of gas  from the intergalactic medium \citep[IGM, e.g., ][]{dekel09, huillier12, hayward12,bouche13}, which drives widespread star formation in galactic disks over Gigayear (Gyr) time-scales \citep[e.g., ][]{daddi10, genzel10, tacchella16, jimenezandrade19}. Such a process of galaxy evolution, known as the ``cold gas accretion mode'' of star formation, differs from the more intense production of stars during occasional  starburst episodes of  ten to few hundred Megayears (Myr) length  \citep[e.g.,][]{daddi10, genzel10, rodighiero11}. This is often due to major/minor mergers providing the energetic and baryonic input to abruptly enhance the star formation rate (SFR) of galaxies \citep[e.g.,][]{narayanan10,huillier12, ellison13}.  Whereas both regimes of star formation  have been { widely explored} out to intermediate redshifts \citep[$z\sim2$;  e.g., ][]{genzel10, wuyts11, magnelli12, magnelli14, daddi15,  elbaz18, jimenezandrade19},  the relative role of the cold gas accretion and merger mode in driving the intense production of stars in galaxies at higher redshifts ($z\gtrsim3$) remains an open issue \citep[e.g.,][]{carilli10, hayward12,  hodge12, hayward18, jimenezandrade18,  tadaki18}.\\

Exploring such early cosmic epochs has been possible thanks to a strong, negative K-correction that makes  high-redshift star-forming galaxies (SFGs) easier to detect at sub-millimeter (sub-mm) wavelengths \citep[e.g., ][]{blain02, casey14}. 
These sub-mm selected galaxies (SMGs)  are, in general, massive star-bursting systems with SFR up to $\sim2000\,M_\odot$\,yr$^{-1}$ and stellar masses ($M_\star$) of $\log(M_\star/M_\odot)\gtrsim 10.5$ 
\citep[e.g., ][]{barger12, smolcic15,gomez-guijarro18,harrington18, jimenezandrade18,lang19}.  SMGs have acquired particular relevance to probe the merger and cold gas accretion mode in the yet unexplored $z\gtrsim3$ regime \citep[e.g., ][]{carilli10, hayward12}. Although the ``canonical'' formation scenario of these massive starbursts  involves major gas-rich mergers \citep[e.g., ][]{tacconi06, tacconi08, engel10, bothwell10, narayanan10, ivison12, bothwell13b},   recent observational (and theoretical) evidence indicates that highly active star-forming disks can also lead to  SMG-like luminosities \citep[e.g., ][]{dave10, hodge12,  hodge16,  narayanan15, hayward18, tadaki18, hodge19}. \\

A heterogeneous SMG population, i.e., secular disks and major mergers, could also explain the diversity of  quiescent massive galaxies at $z\sim2-3$ \citep[e.g., ][]{gobat12, toft12, toft17}. Whereas the  structure and dynamics of most of those quiescent systems seem to be a result of compact, merger-driven SMGs at $z>3$ \citep[][]{toft14, ikarashi15, fudamoto17a, gomez-guijarro18}, the progenitors of  quiescent disk galaxies at $z\sim2$ \citep{newman12, toft17} might have hosted enhanced star formation distributed across a massive rotating disk. \\

Despite the necessity to characterize the properties of $z>3$ massive, star-forming disks, only limited/small samples of such galaxies exist \citep[e.g., ][]{hodge12, debreuck14, jones17, shao17}. For instance, out of the 118 SFGs at  $4<z<6$  in the recent ALPINE [C\,II] survey, no more than 15\% of them are rotating disks \citep{lefevre19}. Among the hundreds of  SMGs across the two square degree COSMOS field \citep[e.g.,][]{casey13, brisbin17},  only AzTEC1 ($z=4.341$),  AzTEC/C159 ($z=4.569$), J1000$+$0234 ($z=4.542$),  and Vd$-$17871 ($z=4.622$)  exhibit {convincing} evidence for gas-dominated rotating disks \citep[][]{jones17, tadaki18}. Consequently, these systems emerge as key laboratories to investigate the  role of cold gas accretion in driving  star formation at $z\sim4.5$, which is the cosmic epoch when the cosmological gas accretion rate onto galaxies is expected to be maximal \citep[e.g.,][]{keres05}.\\  

Here,  we  use high-resolution observations of the Atacama Large Millimeter Array (ALMA) and  NOrthern Extended Millimeter Array (NOEMA) to unveil the redshift and conditions for star formation in AzTEC2. This source is one of the brightest  SMGs in the COSMOS field, which is composed by a massive, star-forming disk and a smaller companion galaxy  at $z = 4.63$.   We use [C\,II] and $^{12}\rm{CO(5\to4)}$ line observations to probe the gas content, star formation efficiency, and gas dynamics of  AzTEC2 within the context of cold gas accretion and merger-driven  star formation in the early Universe. This manuscript is organized as follows. In \S  \ref{aztec2}, we introduce the AzTEC2 source, while in \S \ref{aztec2:sec_observations}, we describe the observations and data reduction. In \S \ref{aztec2:sec_analysis}, we present the analysis and results. The implications of this work are discussed in \S \ref{aztec2:sec_discussion}. We adopt  a flat $\Lambda$CDM cosmology with $h_0 = 0.7$, $\Omega_M = 0.3$, and $\Omega_\Lambda = 0.7$.

\begin{figure*}
	\begin{centering}
		\includegraphics[width=18.1cm]{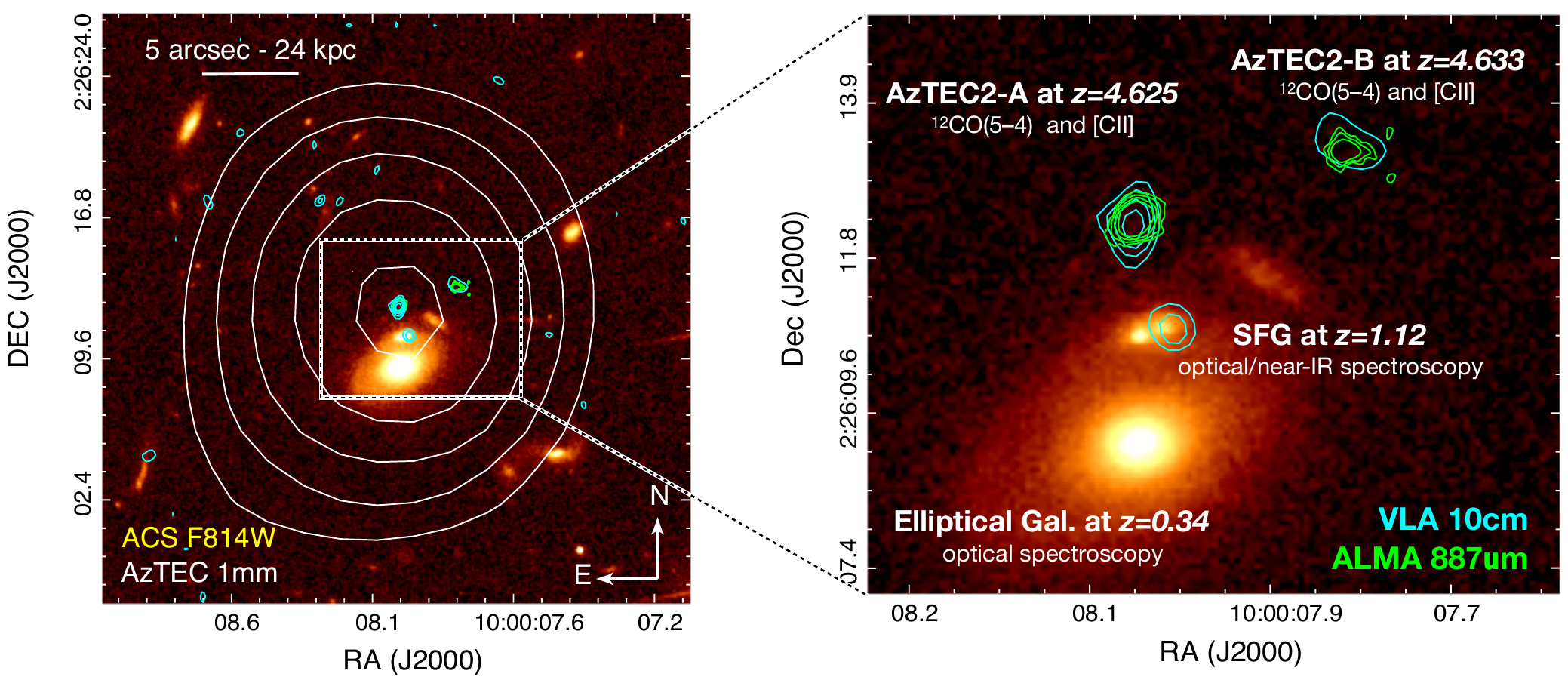}		
		\caption{Multi-wavelength view towards the AzTEC2 source. \emph{Left panel:} AzTEC/JCMT 1.1\,mm \citep[white][]{scott08}, ALMA { Band 7/$887\mu\rm m$ (green; this study)} and VLA Band S/$10\,\rm cm$ \citep[blue; ][]{smolcic17} contours overlaid on top the \textit{HST}/ACS F814W image. All  contour levels are above three times the noise r.m.s.   The zoomed-in image  (\emph{right panel}) shows two foreground sources: a massive elliptical galaxy at  $z=0.34$ and a SFG at $z=1.12$.  ALMA and VLA imaging at sub-arcsec resolution reveal two optically-undetected  components at $z\sim4.63$, labeled as AzTEC2-A and AzTEC2-B. }
		\label{aztec2:panchro_deimos}
	\end{centering}
\end{figure*}

\section{AzTEC2: a bright, multi-component SMG}
\label{aztec2}
AzTEC2 was originally identified in the two  surveys undertaken with the camera AzTEC at 1.1\,mm over an area of $\sim0.5$\,degs$^2$ in the COSMOS field \citep{scott08, aretxaga11}. In the first survey, obtained with the camera AzTEC mounted at the James Clerk Maxwell Telescope (JCMT), it was identified as the second brightest source in the survey: AzTEC2 \citep{scott08}. Later, with the camera AzTEC on the Atacama Submillimeter Telescope Experiment (ASTE), it was identified as AzTEC/C3, i.e.,  the third brightest  source in the survey \citep{aretxaga11}.  This bright SMG was also detected in the deep Herschel/HerMES survey
 (250-500\,$\mu$m maps) and  SCUBA-2 at both 450\,$\mu$m and 850\,$\mu$m \citep{casey13}.
AzTEC2 was recently cataloged as the second brightest SMG  in the  IRAM/GISMO 2\,mm deep survey \citep[over $\sim$250 arcmin$^{2}, $][]{magnelli19} and as the brightest galaxy in the ALMA 2\,mm mosaic in the COSMOS field (with an area of $\sim$155  arcmin$^{2}$; Casey et al. in prep.; Zavala et al. in prep.). \\

Recent imaging with ALMA \citep[][]{brisbin17}  revealed that AzTEC2 is composed of two components  separated by 3\,arcsec (see Fig. \ref{aztec2:panchro_deimos}): namely component AzTEC2-A and AzTEC2-B.  Both sources  were also  detected with signal-to-noise ratio (SNR)\,$>10$  in the 3\,GHz radio continuum imaging with the Very Large Array (VLA) at $0.75$\,arcsec resolution \citep[][]{miettinen17}.   Neither of the two components have a robust optical/near-infrared (IR) counterpart \citep[Fig. \ref{aztec2:panchro_deimos}; ][]{laigle16}, hindering the redshift determination of the AzTEC2 complex. While a spectroscopic redshift solution of $z=1.12$  has been adopted for AzTEC2 in past studies \citep[e.g., ][]{smolcic12, smolcic17c, miettinen15, miettinen17, brisbin17}, recent optical/near-IR spectroscopy   revealed that such a redshift value corresponds to a bright foreground SFG at only 1.5\,arcsec to the south of AzTEC2 \citep[][see Fig. \ref{aztec2:panchro_deimos}]{casey17}. High-resolution observations at sub-mm/mm wavelengths (probing the cold star-forming interstellar medium) are  hence crucial to unambiguously  constrain  the redshift of AzTEC2-A and AzTEC2-B,  as well as to investigate the conditions for star formation in these bright SMGs. \\

\section{Observations and data reduction}
\label{aztec2:sec_observations}

 ALMA Band 7 observations (project 2015.1.00568.S, PI: C. Casey) were conducted on April 23 and  September 1, 2016. The 12\,m main array was used in two different configurations to obtain sub-arcsecond angular resolution without losing sensitivity at larger scales (maximum recoverable angular scale of $\sim4.7$\,arcsec). 
 The spectral setup, originally designed to only detect dust continuum emission,  covered the frequency ranges of 335.5 $-$ 339.5\,GHz and 347.5 $-$ 351.5\,GHz. Data reduction and imaging were performed following the standard steps of the ALMA reduction pipeline scripts with  CASA.   During the imaging process, a Briggs weighting (robust=0.5) was used since it provided a good compromise between  angular resolution and noise. 
We reached a final sensitivity of $1\sigma\simeq$\,1\,mJy\,beam$^{-1}$ 
for a 50\,MHz channel width (corresponding to $\rm \sim 50\,km\,s^{-1}$) and a median restoring beam of $0.23\times0.18$\,arcsec$^2$ (PA=51$^\circ$). These observations allow us to pinpoint the emission from the different sources in this crowded field, and to spatially resolve the emission of AzTEC2-A.\\

A preliminary analysis of the ALMA data revealed a serendipitous line detection at the edge of our spectral windows, which was associated with [C\,II] at $z\sim4.6$ (see details in \S \ref{aztec2:sec_analysis}). To confirm the redshift, we then analyzed observations taken with the Redshift Search Receiver (RSR; \citealt{erickson07}) on the Large Millimeter Telescope.  AzTEC2 was targeted as part of the Early Science Phase observations between 2014 and 2015 with a 32-m antenna  (projects YUNM020 and HUGD024, PIs: M. Yun and D. Hughes, respectively). A total  on-source time of 5\,hrs led to a noise r.m.s of $\approx$\,1.0 mJy\,beam$^{-1}$ per channel, with a spectral resolution of $\sim$\,31\,MHz/100\,km\,s$^{-1}$ and spatial resolution of $\sim25$\,arcsec. Data reduction was performed in a similar way as described in \citet{zavala15,zavala18}. The final spectrum, covering the frequency range 73$-$111\,GHz, revealed a tentative detection of the \hbox{$^{12}$CO$(5\to4)$} line emission at $z\sim4.6$. Although  its low SNR ($\sim$2) prevented us from firmly confirming the redshift, this tentative line detection allowed us to request further observations.\\

Follow-up observations were hence taken with  NOEMA in Band 1  over two tracks on January 26 and 29, 2019 (project W18EU, PI: E.F. Jim\'enez-Andrade). A total observing time of 3.3 hours was reached  using 10 antennas in A-configuration. We used the  PolyFix correlator to cover the frequency range  84.9 -- 92.7\,GHz and 100.2--108.0\,GHz, targeting the  $^{12}$CO($5\to4$) line at $z\sim4.6$. The data reduction was performed with the software \texttt{GILDAS} using the  NOEMA standard pipeline, while imaging was done with the package \texttt{mapping} using natural weighting. The achieved spatial resolution of $1.7\times0.9$ arcsec ($\rm{PA=-163^\circ}$)  suffices to resolve the two components of AzTEC2 separated by $\sim$3\,arcsec (Fig. \ref{aztec2:panchro_deimos}). The spectral data cube was smoothed to a $\sim$34\,MHz resolution (i.e., $\sim100$\,km\,s$^{-1}$), reaching a sensitivity of 0.13\,mJy\,beam$^{-1}$ per channel.\\

\section{Data analysis and Results}
\label{aztec2:sec_analysis}

\subsection{Gravitational lensing magnification}

Although there are no clear indications of  gravitational amplification in our  high-resolution ALMA data (Fig.  \ref{aztec2:submm_specs}), there are two foreground galaxies  which could magnify the emission of both  AzTEC2 components (Fig. \ref{aztec2:panchro_deimos}). We  use the {\tt Visilens} code \citep{spilker16} to estimate the gravitational amplification factor $(\mu)$ as follows. We model each foreground source separately,  adopting a lens mass profile parameterized as an isothermal ellipsoid and   assuming the Einstein mass to be $2.5M_\star$ \citep{auger09}. Given the relatively large offset between the foreground and background galaxies ($\gtrsim 1.5$\,arcsec),  second-order parameters of the lens mass profiles such as shape and ellipticity are found to be not relevant for the analysis. At the position of both AzTEC2-A and AzTEC2-B, the foreground galaxy at $z=1.1$ produces a negligible amplification while using a stellar mass upper limit of $\log(M_\star/M_\odot)=9.6$. On the other hand, the amplification produced by the more massive, elliptical galaxy \citep[$\log(M_\star/M_\odot)=11$;][]{laigle16} at $z=0.34$ is estimated to be $\mu_{\rm A}=1.5$ ($\mu_{\rm B}=1.35$) at the position of AzTEC2-A (AzTEC2-B). We  adopt these magnification factors throughout the rest of the paper.

\subsection{$^{12}$CO(5$\to$4) and [CII] line detections in AzTEC2 at $z=4.6$}
\label{subsec:line_detections}
\begin{figure*}
	\begin{centering}
		\includegraphics[width=14.5cm]{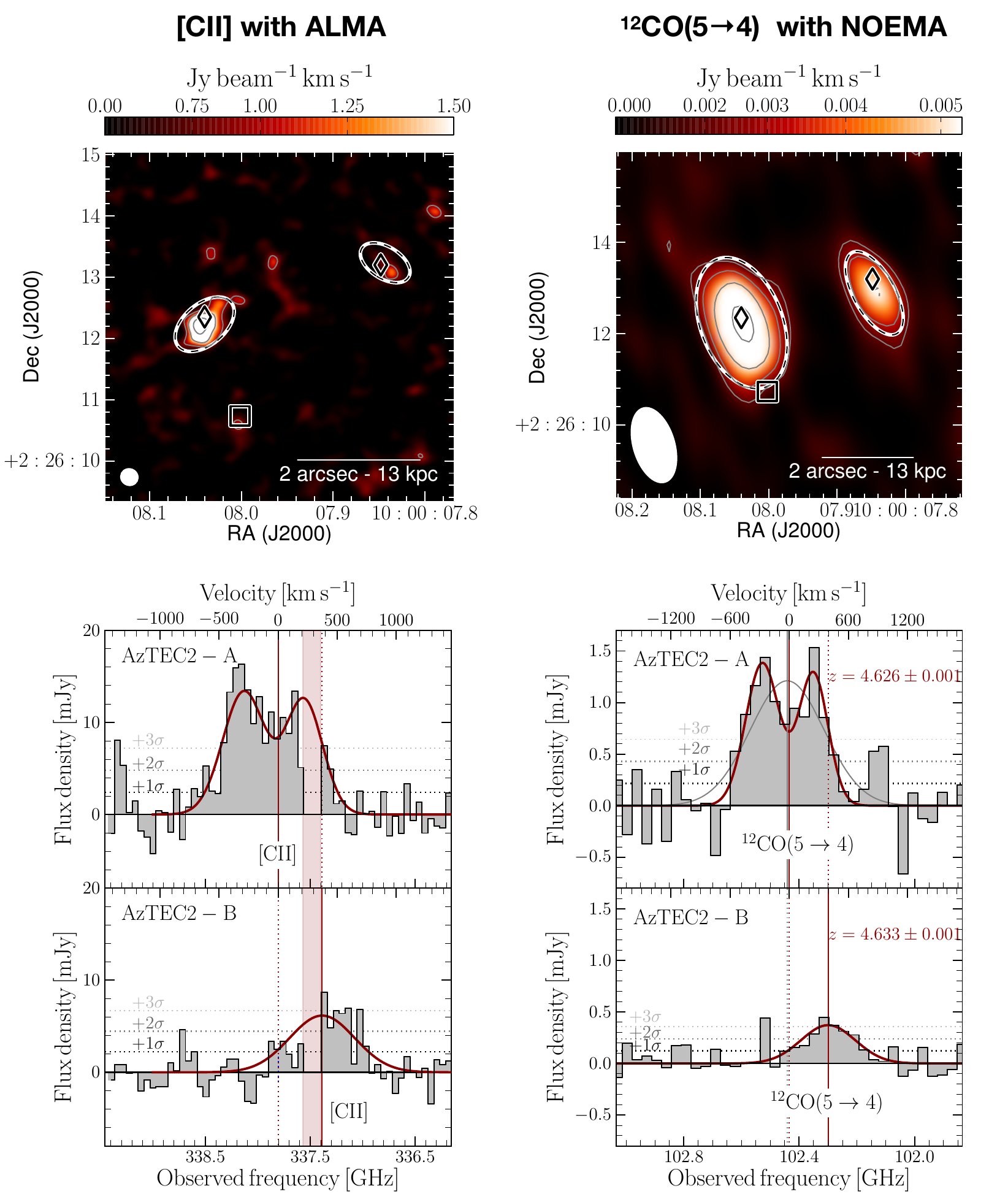}		
		\caption{ \textit{Upper panels:} Velocity-integrated intensity map  ([$-600,+1100]$\,km\,s$^{-1}$)  of the  [C\,II] and $^{12}$CO(5$\to$4)  line emission detected with   ALMA and NOEMA, respectively, of  AzTEC2-A  and  AzTEC2-B. The contours indicate the [3$\sigma$, 5$\sigma$, 8$\sigma$, 13$\sigma$] levels.   The black diamonds mark the position of AzTEC2-A and AzTEC2-B inferred from high-resolution  far-infrared and radio continuum imaging with   ALMA and the   VLA  \citep{brisbin17, smolcic17}. The center of the foreground  SFG at $z=1.1235$ is marked  by the black square.  The synthesized beam is shown in the lower-left corner, while the dashed ellipses illustrate the aperture used to extract the spectra. 
		\textit{Lower panels:}   Spectra of the [C\,II] and $^{12}$CO(5$\to$4) line emission detected with ALMA and NOEMA, respectively, of AzTEC2-A  and  AzTEC2-B. 
		The red solid line represents the model with one (two) Gaussian component(s) that reproduce the [C\,II] and $^{12}$CO(5$\to$4) line profile in AzTEC2-B (AzTEC2-A). 
		The gray line shows a  Gaussian model to fit the $^{12}$CO(5$\to$4) line profile of   AzTEC2-A. The vertical  lines mark the central frequency of the [C\,II] and $^{12}$CO(5$\to$4) line profiles; in the case of AzTEC2-A,  this is  derived from a model with two (in red) and one (in gray) Gaussian component(s). The velocities displayed in the spectra are  relative to the central frequency of the  $^{12}$CO(5$\to$4) line emission in AzTEC2-A.   }
		\label{aztec2:submm_specs}
	\end{centering}
\end{figure*}

The   NOEMA spectrum reveals emission at the position of AzTEC2-A and AzTEC2-B peaking at  $\sim102.5$\,GHz. By collapsing the data cube within the frequency range 102.2 $-$ 102.7\,GHz, which encompasses the full emission line,  we derive the intensity map shown in the upper-right panel of Fig. \ref{aztec2:submm_specs}. A 2D Gaussian fit indicates that AzTEC2-A is resolved by our observations,  with a deconvolved full width at half maximum   of $\sqrt{\mu_{\rm A}}\,{\rm FWHM=}\,1.2\pm0.4$\,arcsec along the major axis.  We  use an aperture that is a factor 1.5 larger than the convolved FWHM of AzTEC2-A  to retrieve  most of the  emission and extract the $^{12}$CO($5\to4$) line spectrum. Since AzTEC2-B  appears as a point-like (unresolved) source,  we integrate emission across a region that equals the size of the synthesized beam to obtain the spectrum (lower-right panel  of Fig. \ref{aztec2:submm_specs}).  We identify a  broad (\rm{FWHM}\,$\sim800$\,km\,s$^{-1}$), double-peaked emission line centered at $102.43\pm0.03$\,GHz associated with AzTEC2-A. Through a least-squares algorithm (Levenberg-Marquardt), we find that a model with two Gaussian components  describes better the line profile than a single Gaussian curve (yielding a reduced $\chi^2$ of 2.0 and 2.6, respectively). We thus adopt the former model and derive an integrated flux density of $\mu_{\rm A} S_{\rm CO(5\to4)} =1070\pm60\,$mJy\,km\,s$^{-1}$. We also identify  an $8.4\sigma$ line detection at the locus of AzTEC2-B  that is centered at $102.30\pm0.02$\,GHz.  A single Gaussian model (reduced $\chi^2=1.2$) leads to an integrated flux density of $\mu_{\rm B} S_{\rm CO(5\to4)} =260\pm30$\,mJy\,km\,s$^{-1}$ (Table \ref{table_aztec2}).  \\

By  averaging line-free channel maps in the $^{12}$CO($5\to4$) data cube, we also detect dust continuum emission from both components {(peak SNR$\gtrsim$10) at  the observed wavelength of $2924\rm \mu m$.  A 2D Gaussian fitting  gives  a total flux density of $0.33\pm0.05\,\rm mJy$ and  $0.09\pm0.02\,\rm mJy$  for AzTEC2-A and AzTEC2-B, correspondingly  (Table \ref{table_aztec2}). Limited by our $1.7\times0.9$\,arcsec  resolution, the dust continuum emission of both components   is not spatially resolved by these NOEMA observations. }  \\

Significant line emission  is detected at $\sim338$\,GHz towards both AzTEC2 components in the   ALMA data cube (upper-left panel of  Fig.  \ref{aztec2:submm_specs}). The velocity-integrated intensity map shows that these sources are resolved. AzTEC2-A, in particular, exhibits  extended emission distributed across $\sim5$ spatial resolution elements.    A 2D Gaussian fit indicates a deconvolved   FWHM  of $\sqrt{\mu_{\rm A}}\,{\rm FWHM}=0.70\pm0.12$\,arcsec along the major axis for AzTEC2-A, and $\sqrt{\mu_{\rm B}}\,{\rm FWHM}=0.53\pm0.18$\,arcsec for AzTEC2-B  (Table \ref{table_aztec2}). To extract the line spectra we use an aperture that is a factor $1.5$ larger than the measured   FWHM, allowing us to recover most of the extended line emission. As illustrated in the lower-left panel of  Fig.  \ref{aztec2:submm_specs}, these   ALMA line detections lie at the edges of the spectral windows used in our observations,  preventing us from recovering/inspecting their total line profiles. There is tentative evidence, however, of a  double-peaked line profile in AzTEC2-A as that observed at $\sim$103\,GHz with  NOEMA. { During the fitting procedure, the relative amplitude and   FWHM  of the two peaks  from this ALMA [CII]  line  detection are fixed to the values of the $^{12}\rm CO(5\to4)$ line emission.}
 The fit indicates a total integrated flux density of $\mu_{\rm A}S_{\rm [C\,II]}=10.9\pm1.2$\,Jy\,km\,s$^{-1}$ and central frequency of $337.8\pm0.2$\,GHz. On the other hand, we use the   FWHM  of our $\sim$103\,GHz  line detection as a prior to model the profile of that at $\sim338$\,GHz in AzTEC2-B (reduced $\chi^2=0.9$), which gives a  total integrated flux density of $\mu_{\rm B}S_{\rm [C\,II]}=4.2\pm0.9$\,Jy\,km\,s$^{-1}$ and central frequency of $337.4\pm0.2$\,GHz. \\
 
 Finally, { we average line-free channel maps in the [C\,II] data cube and detect dust continuum  emission for both AzTEC2 sources  (peak SNR$\gtrsim$13) at the observed wavelength  of $887\,\rm \mu m$  (see Fig. \ref{aztec2:panchro_deimos}).  By fitting a 2D Gaussian  model we derive a total flux density of  $13.3\pm0.5\,\rm mJy$ and  $4.5\pm0.5\,\rm mJy$  for AzTEC2-A and AzTEC2-B, correspondingly  (Table \ref{table_aztec2}). The model also indicates that  the dust continuum emission of both components is spatially resolved by these ALMA  observations, with a deconvolved FWHM  of $\sqrt{\mu}\,{\rm FWHM}\simeq0.35\pm0.03$\,arcsec (Table \ref{table_aztec2}).} \\

By combining the line detections towards AzTEC2-A  at $\sim$102.5\,GHz and 337.5\,GHz, we can unambiguously associate them with $^{12}$CO(5$\to$4) and [C\,II], respectively, leading to a  redshift solution of $z=4.626\pm0.001$. Similarly, for  AzTEC2-B we estimate  a redshift of $z=4.633\pm0.001$, implying a velocity offset of $+375\pm50$\,km\,s$^{-1}$ with respect to AzTEC2-A. These robust line detections and counterpart association rule out the preliminary redshift solution of $z=1.1235$ for AzTEC2 adopted in past studies \citep[e.g., ][]{smolcic12, smolcic17c, miettinen15, brisbin17, miettinen17}.

\subsection{Molecular gas content and star formation rate of AzTEC2}
\label{sec:gascontent}

We use  the $^{12}$CO(5$\to$4) line detections to estimate the  $^{12}$CO(1$\to$0) line luminosity, $L^\prime_{\rm CO(1\to0)}$, and hence infer the molecular gas mass in the AzTEC2 complex. We first consider that due to the higher temperature of  the   CMB at $z=4.6$, the intrinsic value of the $^{12}$CO(5$\to$4) line, $S^{\rm intrinsic}_{\rm CO(5\to4)}$,  is a factor [1/0.8] higher  \citep{dacuhna13} than the one measured from our observations. In using this factor, we assume that both AzTEC2-A and AzTEC2-B harbor a  dense  interstellar medium (ISM) with elevated gas kinetic temperature ($T_{\rm kin} \sim 40\,{\rm K}$) -- as  the majority of  SMGs \citep[e.g., ][]{magnelli12, canameras18}. Therefore,  we find that $\mu\,S^{\rm intrinsic}_{\rm CO(5\to4)}$ is $1340\pm100$\,mJy\,km\,s$^{-1}$  ($325\pm40$\,mJy\,km\,s$^{-1}$) for AzTEC2-A (AzTEC2-B). The corresponding line luminosity is subsequently derived following  \citet[][Sect. 2.4]{carilli13}. We adopt typical SMG-like gas excitation conditions to convert the $^{12}$CO(5$\to$4) line luminosity, $L^\prime_{\rm CO(5\to4)}$,  to  $L^\prime_{\rm CO(1\to0)}$. Then,  $L^\prime_{\rm CO(1\to0)}=[1/0.32]\,\times  L^\prime_{ \rm CO(5\to4)}$   \citep[e.g., ][]{bothwell13, carilli13}, which gives $\mu\, L^\prime_{\rm CO(1\to0)}=12.8\pm2.4\times10^{10}\,{\rm K\,km\,s}^{-1}\,{\rm pc}^{2}$ and $3.1\pm0.7\times10^{10}\,{\rm K\,km\,s}^{-1}\,{\rm pc}^{2}$ for AzTEC2-A and AzTEC2-B, respectively (see Table \ref{table_aztec2}). Finally, the molecular gas mas, $M_{\rm gas}$, can be inferred through the CO-to-H$_2$ ($\alpha_{\rm CO}$) conversion factor: $M_{\rm gas}=\alpha_{\rm CO}L^\prime_{\rm CO(1\to0)}$.  The value of $\alpha_{\rm CO}$ depends on the physical and chemical conditions of the ISM \citep[e.g., ][]{papadopoulos12b, papadopoulos12}. { While  low  values  ($\alpha_{\rm CO}= 0.8 \,M_\odot$\,K$^{-1}$\,km$^{-1}$\,s\,pc$^{-2}$)  are consistent with the turbulent and extreme ISM conditions of, for example, ultra-luminous infrared galaxies \citep[ULIRGs; e.g., ][]{downes98}, higher  vales  ($\alpha_{\rm CO}= 4.3 \,M_\odot$\,K$^{-1}$\,km$^{-1}$\,s\,pc$^{-2}$)   are consistent with a self-gravitating gas configuration as observed in star-forming disks \citep[e.g., ][]{magnelli12}.  Here we adopt a mean $\alpha_{\rm CO}=2.5\,M_\odot$\,K$^{-1}$\,km$^{-1}$\,s\,pc$^{-2}$\footnote{We refer the reader to \S \ref{subsec:discussion} where we discuss in more detail the nature of $\alpha_{\rm CO}$} to better compare  AzTEC2-A and AzTEC2-B with the heterogeneous population of $z\sim4-5$  SMGs, which includes both mergers and star-forming disks \citep[e.g., ][]{hayward18}. }  We find a molecular gas mass  of   $\mu_{\rm A}\,M_{\rm gas}(\alpha_{\rm  CO}/2.5)=3.2\pm0.6\times10^{11}M_{\odot}$ in  \hbox{AzTEC2-A}, and a factor four lower  in AzTEC2-B (see  Table \ref{table_aztec2}). The lensing-corrected gas mass budget  of AzTEC2-A, $M_{\rm gas}(\alpha_{\rm  CO}/2.5)=2.1\pm0.4 \times10^{11}M_{\odot}$, is consistent with the massive gas reservoir of others   $z\sim4-5$ SMGs like AzTEC1, AzTEC3, AzTEC/C159,  J1000$+$0234, and GN20 \citep{schinnerer08,carilli10, riechers10, hodge12, yun15, jimenezandrade18, tadaki18}, { for which the median $M_{\rm gas}(\alpha_{\rm  CO}/2.5)$ is $\sim 2\times10^{11}M_{\odot}$.}  \\

\begin{figure*}
	\begin{centering}
		\includegraphics[width=14.cm]{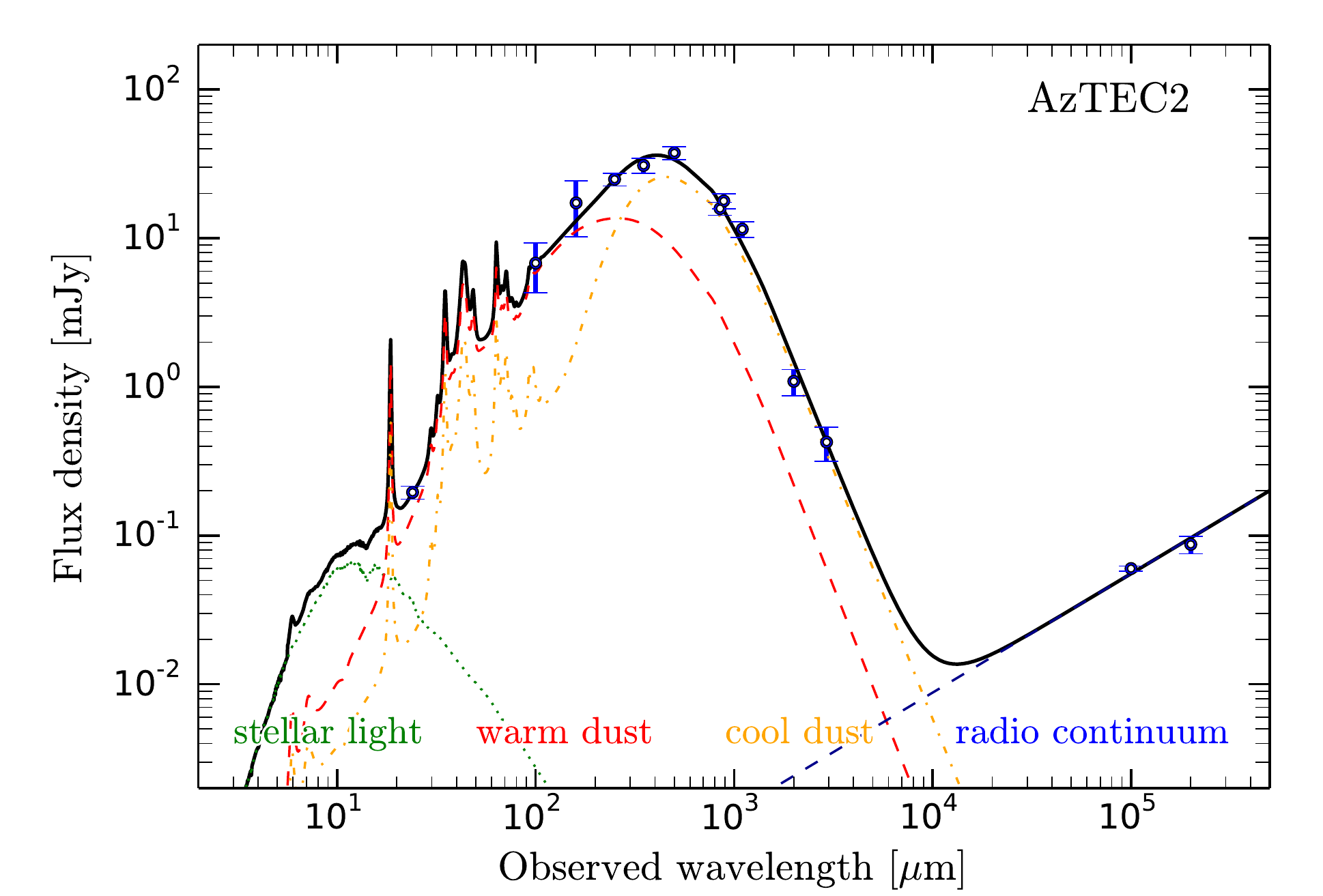}
		\caption{Broadband SED of AzTEC2. The monochromatic  flux densities (blue  circles) used in the fit correspond to the total emission from both components (see values in Table \ref{table_aztec2}). The best fit model is given by the black thick line. The dashed red (dash-doted  orange)  line shows the   model of the warm (cool) dust emission. The stellar emission is shown by the dotted green line, while the  radio continuum emission is represented by the dashed blue line.} 
		\label{aztec2:fir_sed}
	\end{centering}
\end{figure*}

 We derive the infrared luminosity $(L_{\rm IR}$, integrated over the wavelength range of $8-1000\,\mu$m) and dust mass $(M_{\rm dust})$ of the AzTEC2 complex  by fitting its mid-IR--to--mm SED. This is done by { following the SED fitting procedure presented by \citet{liu18}}, and by combining our dust continuum measurements at 887\,$\mu$m and 2.92\,mm  with information from the  COSMOS photometric catalog compiled by \citet{jin18} {and} \citet{liu18}. In the case of AzTEC2, this catalog includes photometric measurements at (see Table \ref{table_aztec2}): $\rm 24\,\mu m$ \citep{lefloch09}, $\rm 100\,\mu m$,  $\rm 160\,\mu m$ \citep{lutz11}, $\rm 250\,\mu m$, $\rm 350\,\mu m$,  $\rm 500\,\mu m$ \citep{oliver12}, $\rm 850\,\mu m$ \citep{geach16}, $1.1\,\rm mm$ \citep{aretxaga11}, 10\,cm \citep{schinnerer10} and 21\,cm \citep{smolcic17}.  The  photometric measurement at 2\,mm recently obtained with the GISMO-2 bolometer camera \citep{magnelli19} is also included in the analysis. \\

 Since most photometric data towards the AzTEC2 complex do not have sufficient resolution   to  { deblend the emission of} AzTEC2-A and AzTEC2-B (Table \ref{table_aztec2}),    we use the combined monochromatic flux densities of both  components even if high-resolution observations are available.   We then fit the SED  with five components: a stellar template from the \citet{bruzual03} models, an active galactic nuclei (AGN) template from \cite{mullaney11}, warm and cold dust templates from \cite{draine07b}. The fifth component is  a radio power law  tailored to the dust IR luminosity with $q_{\mathrm{IR}}=2.4$, where $q_{\mathrm{IR}}$ is the median value for the ratio between far-infrared and radio luminosity of SFGs \citep[e.g., ][]{magnelli15, delhaize17}. The fitting is performed through  Monte Carlo sampling (with $N=15000$, following \citet{liu18}), from which  the $\chi^2$ distribution and  uncertainties are obtained. Our analysis indicates that a model with no AGN component provides the best fit to our data points, albeit more photometric information  is needed to confirm the (apparently) negligible AGN activity in AzTEC2. Finally, we derive a total infrared luminosity of $\log(\mu L_{\rm IR}/L_{\odot})= 13.59\pm0.02$ and dust mass of $\log(\mu M_{\rm dust}/M_{\odot})=9.64\pm0.10$ for the AzTEC2 complex.\\

To disentangle the contribution  of AzTEC2-A and AzTEC2-B to the   total $L_{\rm IR}$ and $M_{\rm dust}$, we use our high-resolution photometric data  at $887\mu$m and $2924\mu$m.  These observations independently trace the peak and Rayleigh-Jeans regime of the SED of both components (Table \ref{table_aztec2}), allowing us to infer the contribution of AzTEC2-A and AzTEC2-B to the total IR SED.  
{
	Then, since the ratio between the IR flux density of the two components at $887\mu$m and $2924\mu$m  is $S^{\rm IR}_{\rm AzTEC2-A}/S^{\rm IR}_{\rm AzTEC2-B} \simeq 3$, we estimate  $\log(\mu_{\rm A}L_{\rm IR}/L_{\odot})=13.46\pm0.02$ and $\log(\mu_{\rm A}M_{\rm dust}/M_{\rm \odot})=9.51\pm0.10$ for AzTEC2-A. For AzTEC2-B we derive $\log(\mu_{\rm B}L_{\rm IR}/L_{\odot})=12.98\pm0.02$ and $\log(\mu_{\rm B}M_{\rm dust}/M_{\rm \odot})=9.03\pm 0.10$.     \\

We note that the above reasoning assumes that the  intrinsic SED of AzTEC2-A and AzTEC2-B are similar. While this is supported by the consistent flux density ratios at $887\mu$m and $2924\mu$m, the  properties that have been inferred from  the scaled SEDs are subject to a high degree of uncertainty. Therefore, { to add an independent constraint on the  $L_{\rm IR}$ of AzTEC2-A and AzTEC2-B that is not affected by source blending, we use their [C\,II] line luminosity $(L_{[\rm C\,II]})$ to estimate $L_{\rm IR}$ via the empirical  $L_{[\rm C\,II]}/L_{\rm IR}$ luminosity ratio \citep[e.g., ][]{maiolino09, lagache18}.  We assume that the physical properties of AzTEC2-A and AzTEC2-B  are  similar to the ones of bright SMGs at similar redshifts ($4<z<5$),  for which  $L_{\rm [C\,II]}/L_{\rm IR}=7^{+4}_{-2} \times 10^{-4}$  \citep[see compilation in Table B.1 of ][]{lagache18}. With $\log(\mu_{\rm A}L_{[\rm C\,II]}/L_\odot)=10.49\pm0.05$, we estimate $\log(\mu_{\rm A}L_{\rm IR}/L_{\odot})=13.48\pm0.14$  for AzTEC2-A. Likewise, for AzTEC2-B we find $\log(\mu_{\rm B}L_{[\rm C\,II]}/L_\odot)=10.07\pm0.10$ and $\log(\mu_{\rm B}L_{\rm IR}/L_{\odot})=13.07\pm0.17$.  Although here we neglect possible differences in the  $L_{[\rm C\,II]}/L_{\rm IR}$  ratio of  AzTEC2-A and AzTEC2-B arising from distinct physical conditions of the ISM  \citep[e.g.,  ultraviolet radiation field and/or metal enrichment; ][]{katz17, rybak19}, these new and independent $L_{\rm IR}$ estimates corroborate  those derived via SED fitting. }\\

We infer the  SFR  of both components following the calibration from \citet{kennicutt98}: SFR[$M_\odot$\,yr$^{-1}$]$=10^{-10}L_{\rm IR}\,[{L_\odot}]$. Assuming a Chabrier  Initial Mass Function, we derive  $\mu_{\rm A}\,{\rm SFR}=2880\pm 140\,M_\odot$\,yr$^{-1}$ for AzTEC2-A, while for AzTEC2-B we find  $\mu_{\rm B}\,{\rm SFR}=960\pm 45\,M_\odot$\,yr$^{-1}$.  The  lensing-corrected  SFR of  AzTEC2-B ($710\pm 35\,M_\odot$\,yr$^{-1}$) is consistent with the average for
 SMGs at similar redshifts \citep[$\sim800\,M_\odot$\,yr$^{-1}$; e.g., ][]{smolcic15, gomez-guijarro18, jimenezandrade18, magnelli19}.  The  extreme IR-based SFR of AzTEC2-A (SFR=$1920\pm 100\,M_\odot$\,yr$^{-1}$) is  comparable to those of the massive, star-forming disks GN20 and AzTEC1 at $z\sim4.5$ \citep{carilli10, magnelli12, tadaki18}. \\

 On the other hand, we use our $M_{\rm dust}$ estimates to infer the molecular gas mass of AzTEC2-A and AzTEC2-B. We assume that these SMGs have solar metallicity and, consequently, that they harbor a gas-to-dust ratio of $\delta_{\rm GDR}\sim 100$ (following the $\delta_{\rm GDR}$ -- metallicity relation derived by \citet[][]{leroy11}). Then, $\log(\mu_{\rm A} M_{\rm gas}/M_\odot) \simeq \log(\mu\delta_{\rm GDR} M_{\rm dust}/M_\odot)=11.5\pm0.1$ for AzTEC2-A and $\log(\mu_{\rm B} M_{\rm gas}/M_\odot)=11.0\pm0.1$ for AzTEC2-B. These values are in good agreement with the  $M_{\rm gas}$ estimates derived from our $^{12}$CO$(5\to4)$ line observations assuming a mean $\alpha_{\rm CO}$ of 2.5\,$M_\odot\,{\rm K}^{-1}\,{\rm km}^{-1}\,{\rm s}\,{\rm  pc}^{-2}$, which are $\log(\mu\,M_{\rm gas}/M_\odot)=11.50\pm0.05$ and $10.90\pm0.05$ for AzTEC2-A and AzTEC2-B, correspondingly.

\subsection{Mode of star formation  in AzTEC2}
\label{sec:modesf}

The $L_{\rm IR}/L^{\prime}_{\rm CO(1\to0)}$ ratio  gives an indication of how efficient  the production of stars in galaxies is for a given molecular gas reservoir. We estimate a $L_{\rm IR}/L^\prime_{\rm CO(1\to0)}$ ratio of  $220\pm50$ \,$L_\odot({\rm K\,km\,s}^{-1}{\rm pc}^{2})^{-1}$ for AzTEC2-A and  $300\pm85\,L_\odot({\rm K\,km\,s}^{-1}{\rm pc}^{2})^{-1}$ for  AzTEC2-B (see Table \ref{table_aztec2}).  The  star formation efficiency of AzTEC2-A  is larger than 
those of nearby and $z\sim 2$ star-forming disks \citep[20$-$100\,$L_\odot({\rm K\,km\,s}^{-1}{\rm pc}^{2})^{-1}$; ][]{daddi10, genzel10}, but it is in agreement with the $L_{\rm IR}/L^\prime_{\rm CO(1\to0)}$ ratio of  $z\sim4.5$ star-forming disks like GN20 and  AzTEC/C159 \citep[180$-$220\,$L_\odot({\rm K\,km\,s}^{-1}{\rm pc}^{2})^{-1}$; ][]{hodge12, jimenezandrade18}. \\

To better compare the star formation efficiency of AzTEC2-A and AzTEC2-B with respect to the overall SFG's population,  in Fig. \ref{fig:aztec2_lir-lco}  we present the $L_{\rm FIR}-L^\prime_{\rm CO(1\to0)}$ relation for star-forming disks and merger-driven starbursts derived by \citet{genzel10}. For this exercise,  we estimate the far-IR (FIR) luminosity ($L_{\rm FIR}$) of the AzTEC2 complex by integrating the total IR SED (Fig. \ref{aztec2:fir_sed}) over the wavelength range 42.5 -- 122.5\,$\mu$m \citep[following ][]{helou85}. Then, the $L_{\rm FIR}$ of AzTEC2-A and AzTEC2-B are inferred from the relative ratio of their dust-continuum flux density at $887\mu$m and $2924\mu$m (as done for $L_{\rm IR}$ in \S \ref{sec:gascontent}). We estimate $\log(\mu_{\rm A}L_{\rm FIR}/L_\odot)=13.08\pm0.02$  for AzTEC2-A, and $\log(\mu_{\rm B}L_{\rm FIR}/L_\odot)=12.60\pm0.02$ for AzTEC2-B (see Table \ref{table_aztec2}). Combining these values with our $L^\prime_{\rm CO(1\to0)}$ estimates, we plot the locus of 
AzTEC2-A and AzTEC2-B in the $L_{\rm FIR}-L^\prime_{\rm CO(1\to0)}$ plane. We also include  compilations of nearby normal and starburst galaxies \citep{solomon97, gao04}, $z\sim1.5$ star-forming disks \citep{daddi10, geach11, magnelli12}, SMGs \citep{bothwell13}, and massive star-forming disks at $z\sim4.5$ \citep{carilli10, hodge12, jones17, jimenezandrade18, tadaki18}. This comparison indicates that most of the reported  SMGs, like AzTEC2-B,  lie on/above the empirical $L_{\rm FIR}-L^\prime_{\rm CO(1\to0)}$ relation for  mergers. Despite being at the high-end of the $L_{\rm FIR}-L^\prime_{\rm CO(1\to0)}$ plane,   AzTEC2-A (as well as GN20 and AzTEC1)  approaches  the relation  of  normal,  star-forming disk galaxies \citep[see Fig. \ref{fig:aztec2_lir-lco}; e.g., ][]{genzel10}.  \\

Another indicator of star formation efficiency is the $L_{\rm IR}/L^{\prime}_{\rm CO(5\to4)}$ ratio \citep[e.g., ][]{daddi15}, which 
 traces the dense,  warm molecular gas ($n>10^{4}\;$cm$^{-3}$) that is closely linked to massive star formation. We find that  AzTEC2-A (AzTEC2-B) exhibits a ratio that is 1.6 (2.5) larger than  local star-forming and $z\sim1.5$ main sequence (MS) galaxies \citep[e.g., ][]{liu15}. 
This suggests that  AzTEC2-A  consumes its star-forming gas faster than secular star-forming disks, but at a more moderate rate than typical SMGs like AzTEC2-B. \\

Given their available gas reservoir, AzTEC2-A  (AzTEC2-B) will be able to sustain their current   SFR for a period  of $\sim$110\,Myr (80\,Myr)   (assuming a mean $\alpha_{\rm CO}=2.5\,M_\odot$\,K$^{-1}$\,km$^{-1}$\,s\,pc$^{-2}$). The  gas depletion time-scale ($\tau_{\rm gas}$)  of AzTEC2-A, in particular,
exhibit a mild excess with respect to the  average for  SMGs at similar redshifts \citep[$\tau_{\rm gas}  \sim 45$\,Myr; ][]{aravena16}, but resembles the one of the massive, star-forming disk galaxies GN20, AzTEC/C159 and AzTEC1 at $z\sim4.5$ \citep{hodge12, jimenezandrade18, tadaki18}.\\

\begin{figure}
	\begin{centering}
		\includegraphics[width=8.5cm]{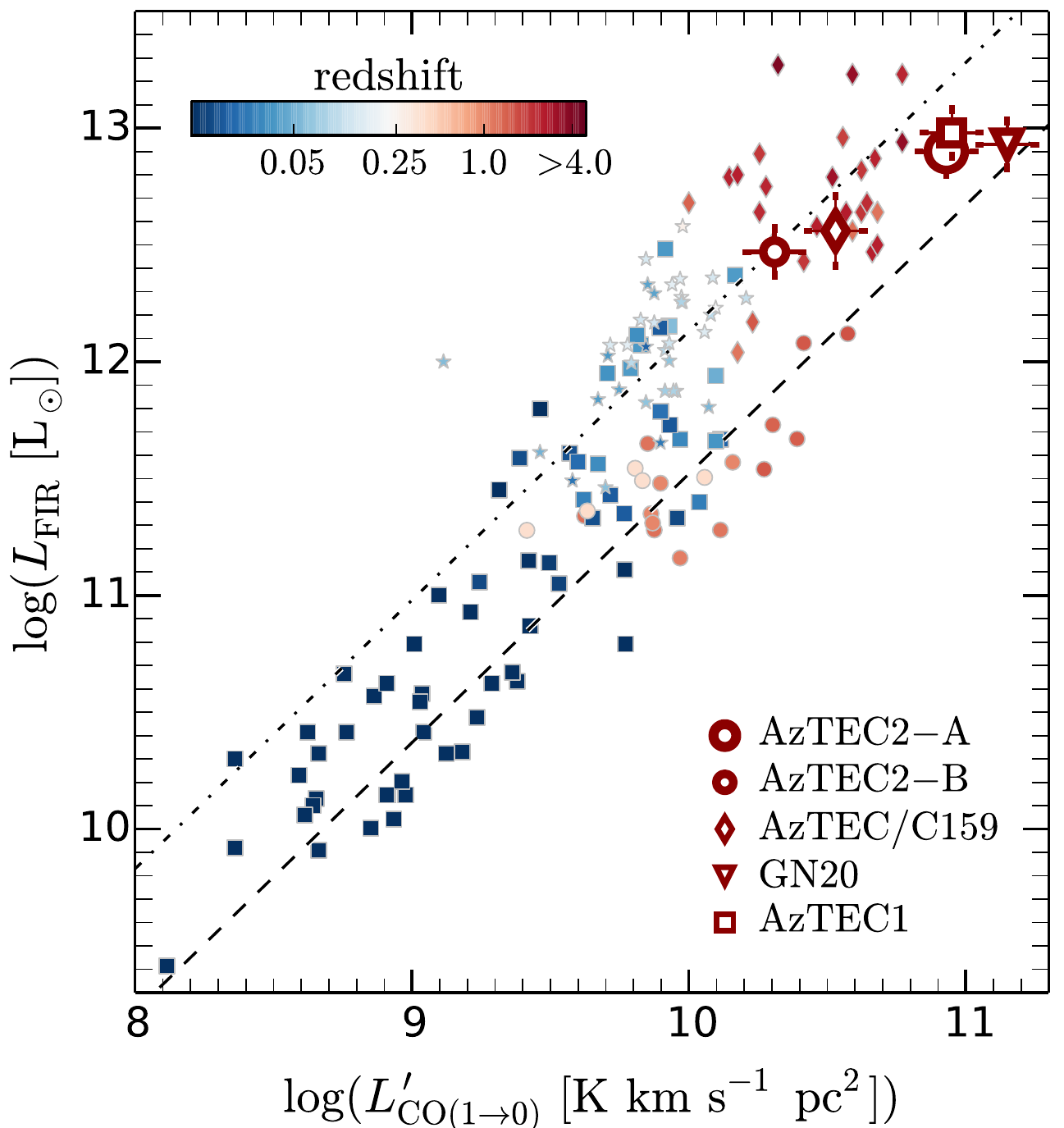}		
		\caption{FIR luminosity  as a function of $^{12}$CO(1$\to$0) line luminosity for local and high-redshift  SFGs.  The squares represent nearby normal and starburst galaxies reported by \citet{gao04}, while stars correspond to low-redshift  ULIRGs in the sample of  \citet{solomon97}.  The  circles show $z\sim1.5$ star-forming disks presented in \citet{daddi10, geach11} and \citet{magnelli12}.  The diamonds represent the parameter space covered by the  SMGs  reported in  \citet{bothwell13}. Large symbols correspond to the SMGs studied here and the  massive, rotating disk galaxies at $z\sim4.5$: GN20, AzTEC1, and AzTEC/C159 \citep{carilli10, hodge12, jones17, jimenezandrade18, tadaki18}. In the case of AzTEC2-A and AzTEC2-B, we use lensing-corrected luminosities.   If only $L_{\rm IR}$ measurements are available in the literature, we convert  $L_{\rm IR}$ into $L_{\rm FIR}$ by considering that $\left< \log(L_{\rm IR})\right>=0.3+\left< \log(L_{\rm FIR})\right>$ \citep[e.g., ][]{delhaize17}. The dashed and dotted lines show the best-fitting relation  for  MS and starburst galaxies, respectively, reported by  \citet{genzel10}. }
		\label{fig:aztec2_lir-lco}
	\end{centering}
\end{figure}

Overall,  the properties of AzTEC2-A and AzTEC2-B are   consistent with the  intense star formation activity observed  in  bright SMGs  at $z\sim4$ \citep{schinnerer08,riechers10, yun15, jimenezandrade18, tadaki18}.  AzTEC2-A  resembles --to some extent--  the properties of  massive, star-forming disks at lower  and similar redshifts \citep[e.g., ][]{daddi10, genzel10, hodge12}, which  form stars through the cold gas accretion mode of star formation.

\begin{figure*}
	\begin{centering}
		\includegraphics[width=18cm]{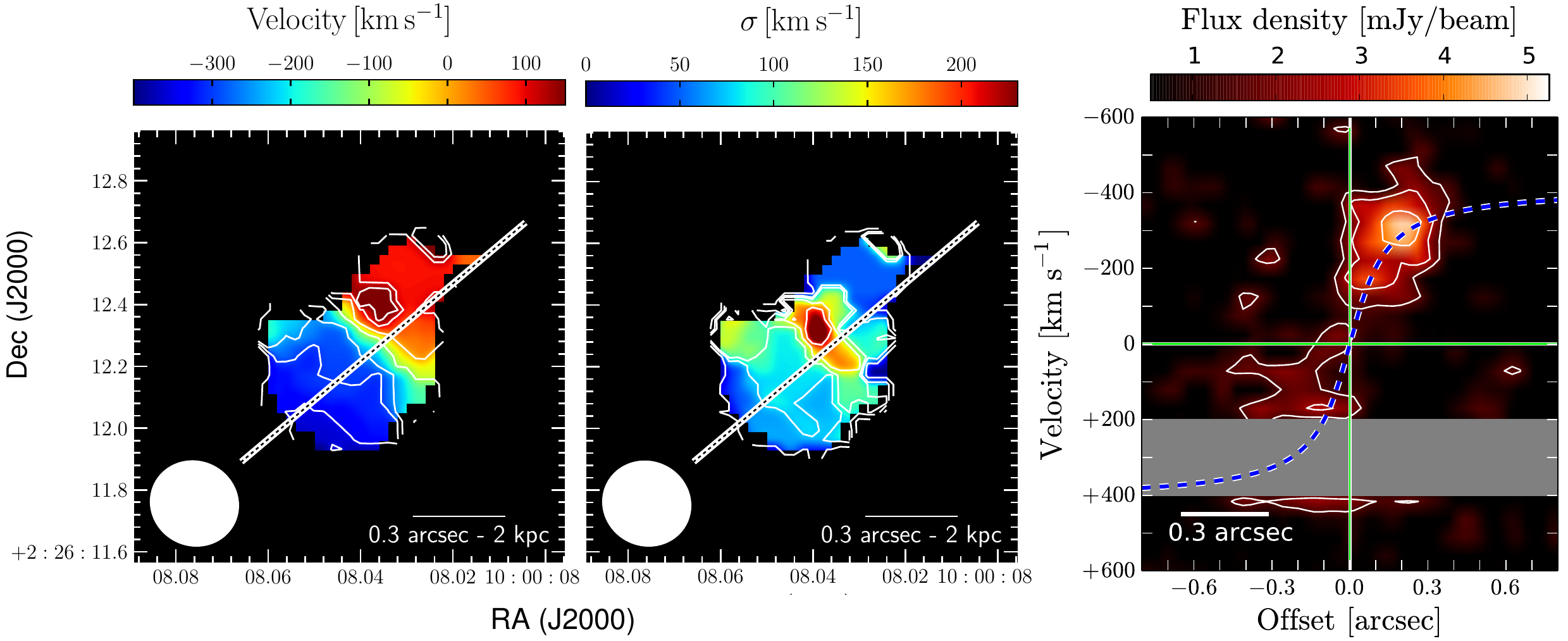}		
		\caption{Velocity field (\textit{left panel}) and velocity dispersion   (\textit{central panel}) of the gas in AzTEC2-A ($z=4.626$), derived from [CII] line observations with  ALMA. Note that the velocity channels above 200\,km\,s$^{-1}$ are not available in the data set (see Fig. \ref{aztec2:submm_specs}). The contour levels are at $[-350, -300, -250, -100, 0, 100, 150]$\,km\,s$^{-1}$  and   $[25, 75,100,150, 200]$\,km\,s$^{-1}$ for the velocity field and velocity dispersion maps, respectively. The synthesized beam is shown in the lower-left corner.  The pv diagram (\textit{right panel}) has been extracted using a 0.3\,arcsec width aperture along the galaxy's major-axis (position angle of $-40^\circ$; {black dashed line}). The blue-dashed line is a simple arctan model to describe the rotation curve of AzTEC2-A. The contour levels are at 3, 5 and 8 times the rms noise. The gray shaded region shows the velocity range that is not available in the current data set.  The horizontal bar shows the  major axis of the synthesized beam.}
		\label{fig:velocitymaps}
	\end{centering}
\end{figure*}

\subsection{A rapidly rotating, massive disk in AzTEC2-A}
\label{subsec:dynamics}

{ Predictions from numerical simulations \citep[e.g.,][]{kohandel19} and observations  of high-redshift galaxies \citep[e.g.,][]{jones17} have suggested that a double-peaked  [C\,II] line profile,  like that  of AzTEC2-A (Fig. \ref{aztec2:submm_specs}), can be consistent with a rotating disk galaxy.} We then use the high-resolution [C\,II] line observations to explore the kinematics of AzTEC2-A, for which the high  SNR of the detection  enables us to derive the  velocity field and   velocity dispersion of the gas.  { As observed in  Fig. \ref{fig:velocitymaps}, AzTEC2-A exhibits a  smooth velocity gradient that appears to be consistent with rotationally dominated kinematics. }  To parameterize  its motion,  we derive the position-velocity (pv) diagram (Fig. \ref{fig:velocitymaps}) along the major-axis  of the [C\,II] line emission using a 0.3\,arcsec width aperture. \\

{ There exist several parameterizations (or models) that can describe the rotation curves of galaxies, including the basic 2-parameter arctan function, the more elaborate ``multi-parameter function'' \citep{courteau97}, and  the ``universal rotation curve''  \citep{persic96}. The empirically motivated arctan model is given by $v(r)=(2/\pi) v_{\rm asym} \arctan(r/r_{\rm t})$ \citep[e.g., ][]{courteau97, willick99}, where  $v_{\rm asym} $ is the asymptotic rotational velocity and $r_{\rm t}$ is the  transition radius between the rising and flat part of the rotation curve. With only two free parameters, the arctan function provides an adequate description to rotation curves of low- and high-redshift galaxies  \citep{courteau97, willick99, miller11, swinbank12, debreuck14, drew18}, while avoiding strong covariances between the fitted parameters of more complex models  \citep{courteau97}. \\

We thus adopt the arctan model to fit the rotation curve of AzTEC2-A. As observed in Fig. \ref{fig:velocitymaps}, this function offers a reasonable description of the observed portion of the PV diagram.} Our fit, limited by the lack of velocity channels above $+200$\,km\,s$^{-1}$, suggests an asymptotic velocity of $415\pm70$\,km\,s$^{-1}$ and $\mu_{\rm A}\,r_{\rm t}=0.7\pm0.1$\,kpc. To derive the intrinsic (de-projected)  rotational velocity, $v_{\rm rot}^{\rm int}=v_{\rm rot}/\sin(i)$, we need to { infer the  disk inclination ($i$). This can be derived from the apparent ellipticity of the galaxy: i.e.,  $i=\arcsin({\rm FWHM}_{\rm minor}/{\rm FWHM}_{\rm major})$, where the respective  FWHM  values can be derived from the  surface brightness distribution of the [C\,II] line. } However, as observed in Fig. \ref{fig:velocitymaps}, the absence of velocity channels above $+200$\,km\,s$^{-1}$ biases the spatial distribution of the [C\,II] line emission. { Assuming a co-spatial distribution of interstellar dust and gas, the inclination of the [C\,II] disk can be approximated from the ellipticity of the dust continuum emission revealed by our high-resolution ALMA observations  (\S \ref{subsec:line_detections}). We then derive  $i=\arcsin(0.23\pm0.02/0.36\pm0.02)=39\pm3^\circ$, which leads to an intrinsic rotational velocity of  $v_{\rm rot}^{\rm int}=v_{\rm asym}=660\pm130$\,km\,s$^{-1}$.} Such a high rotational speed is consistent with the one of the rapidly spinning, star-forming disks GN20, AzTEC/C159 and J1000+0234 at $z\sim4.5$ \citep[$\gtrsim 500$\,km\,s$^{-1}$;][]{carilli10, hodge12, jones17}.   \\

To evaluate the rotational-to-dispersion support ($v_{\rm rot}/\sigma)$ of the disk, we inspect the line-of-sight velocity dispersion ($\sigma$) map in Fig. \ref{fig:velocitymaps}. We anticipate that at the innermost region of the galaxy, the measured $\sigma$  is highly  enhanced by ``beam smearing''\footnote{At the innermost region of galaxies, the measured line width is boosted by large-scale motions occurring within the region traced by a relatively coarse (finite)  Point Spread Function (PSF).} \citep[e.g.,][]{davies11, scott16}. Since such a contribution is expected to be modest at the outermost radii, we adopt $100$\,km\,s$^{-1}$ (from the contour levels in the map, Fig. \ref{fig:velocitymaps}) as an upper limit for the intrinsic velocity dispersion of the gas in the disk. Then, we derive $v_{\rm rot}/\sigma\gtrsim5$,    indicating that  AzTEC2-A  is rather an unperturbed, rotation-dominated  disk that   resembles the $v_{\rm rot}/\sigma$ ratio of more evolved disk galaxies at $z\sim1$  \citep[e.g., ][and references therein]{diteodoro16}. This finding, therefore,   provides more evidence of kinematically mature disks that can be found at  even $z\sim4.5$, such as GN20 \citep{hodge12},   ALESS 73.1 \citep{debreuck14}, AzTEC/C159, J1000$+$0234, \citep{jones17, jimenezandrade18} and AzTEC1  \citep{tadaki18}. \\

Finally, by assuming that the  kinematics of the disk is mainly dominated  by the gravitational potential of AzTEC2-A,  its dynamical mass ($M_{\rm dyn}$)  can be estimated through  the relation: $M_{\rm dyn}\sin^2(i)=Rv(R)^2/G$, where $v(R)$ is the rotation velocity at radius $R$ and  $G$ is the  gravitational constant.  Using  $R=0.5$\,arcsec that encompasses the full extent of the [C\,II] line emission, and that equals the size of the aperture used to obtain the spectrum, we find $M_{\rm dyn}(i=39^\circ)= 2.6^{+1.2}_{-0.9}\times10^{11}{M}_\odot$.  This mass budget  roughly agrees with that expected  for the molecular gas mass ($[0.7, 3.7]\times10^{11}\,M_\odot$), indicating that AzTEC2-A  is  a massive gas-rich disk   possibly assembled through an enhanced accretion of gas from the cosmic web \citep[e.g., ][]{bournaud09, dekel09b, dekel09, romano-diaz14}. A  post-merger scenario, however,  can not be  excluded, given that disks could survive or re-form rather quickly after a gas-rich merger   \citep[e.g.,][]{springel05,hammer09, hopkins09}.  
Simulated disk galaxies that  formed via gas-rich mergers can resemble the observed properties (kinematics, SFR, gas surface density) of  $z\sim2$ disks \citep{robertson08}. Furthermore, a coarse PSF (like ours, $\gtrsim 1$\,kpc) diminishes the contrast between disturbed kinematics and rotation-dominated disks \citep{hung16}. Deeper [CII] line observations with  sub-kpc scale resolution are thus needed to accurately derive the kinematic properties of AzTEC2-A and, hence, to isolate the formation scenario of its rotating gas disk. \\

\section{Implications  for galaxy evolution at high redshift}
\label{aztec2:sec_discussion}
The redshift identification and subsequent dynamical characterization of  AzTEC2-A  add new evidence on the existence of massive vigorously star-forming disks in the early Universe.  By including AzTEC2-A, the sample of SMGs at  $4<z<5$ with robust evidence of rotation has now increased to  seven sources:  AzTEC2-A, AzTEC/C159,  AzTEC1, Vd-17871,  J1000$+$0234, ALESS 73.1, and GN20 \citep[][]{carilli10, hodge12, debreuck14, jones17, tadaki18}. The former galaxies represent more than half the population of spectroscopically confirmed $z > 4$ SMGs within the two square degrees of the COSMOS field \citep{smolcic12, smolcic15}. Here, we discuss the implications of these findings within the context of cold gas accretion and merger-driven star formation in massive,  high-redshift galaxies. \\

\subsection{A heterogeneous SMG population}

The enhanced production of stars in SMGs has been largely attributed to gas-rich galaxy mergers \citep[e.g., ][]{tacconi08, younger10, engel10, riechers11, Iono16}, which is compatible with the merger-driven starbursts in local ULIRGs \citep[e.g.,][]{sanders96}. {In the case of AzTEC2,  and as observed in Fig. \ref{aztec2:panchro_deimos}, our current  ALMA dust continuum imaging does not reveal clear signs of disturbance/interaction between AzTEC2-A and AzTEC2-B (e.g., strong tidal tails and/or bridges).  Although this could be a result of the surface brightness limit of these observations, the clear spatial}  separation between AzTEC2-A and AzTEC2-B  ($\sim$20\,kpc) and the relative velocity offset of 350\,km\,s$^{-1}$ suggest that  these galaxies undergo a pre-coalescence (first approach) phase \citep[e.g.,][]{calderoncastillo19}. { This might indicate that the vigorous SFR in AzTEC2-A is not  dominated by merging activity. Instead,  }the gas velocity fields,  $L_{\rm FIR}/L^\prime_{\rm CO(1\to0)}$, and $L_{\rm IR}/L^\prime_{\rm CO(5\to4)}$  ratio   point towards a smoother mode of star formation that drives a massive, star-forming  disk \citep[e.g., ][]{dekel09b, dekel09, carilli10, hodge12, romano-diaz14}.  Certainly, the  properties of AzTEC2 resemble those of the well-characterized star-forming disk GN20 at similar redshift \citep{carilli10, hodge12}.\\

These results strengthen the scenario in which  single-dish selected SMGs are a heterogeneous population  \citep[e.g.,][]{hayward11, hayward13b}, including  major mergers \citep[e.g., ][]{engel10, riechers14},  isolated disk galaxies and pairs of (likely infalling) galaxies  that are blended into a single sub-mm source as observed in AzTEC2. As  discussed by \citet{hayward11}, this heterogeneity is linked to the SMG selection function. Since sub-mm surveys lead to flux ($\propto$SFR) limited samples of galaxies \citep[e.g.,][]{scott08, aretxaga11}, at high redshift ($z\gtrsim3$), only SFGs harboring a SFR$\gtrsim 300\,M_\odot\,\text{yr}^{-1}$ can be selected with typical single-dish surveys \citep[e.g.,][]{magnelli19}. Therefore,  this selection function  tends to identify  the extreme and massive end of the SFG population at high redshifts, including both merger-driven and massive star-forming disks that sustain vigorous star formation activity leading to SMG-like IR luminosities \citep[e.g.,][]{hayward12}.

\subsection{The cold gas accretion and merger mode of star formation at $z>3$}
\label{subsec:discussion}

The evidence of rotation-dominated, gas-rich, star-forming disks   at $z>4$ \citep[e.g., ][]{carilli10, hodge12, debreuck14, jones17, tadaki18} indicates that high-redshift SMGs are not only merging, strongly perturbed systems \citep{tacconi08, younger10, engel10, riechers11, menendez-delmestre14, chen15, jones17, chang18}. High-resolution (0.07 arcsec) dust continuum imaging with ALMA  has even shown evidence for spiral arms, bars, and rings in $z\gtrsim2$ SMGs  \citep[][]{hodge19}. 
The remaining question is what are the mechanisms  leading to the intense production of stars in such  massive, star-forming disks in the early Universe. We thus infer the locus of AzTEC2-A (and AzTEC2-B) in the Kennicutt-Schmidt  plane (Fig. \ref{fig:aztec2_ks}) and use them  as an observational diagnostic for constraining the global conditions for star formation in these systems. \\

\begin{figure}
	\begin{centering}
		\includegraphics[width=8.85cm]{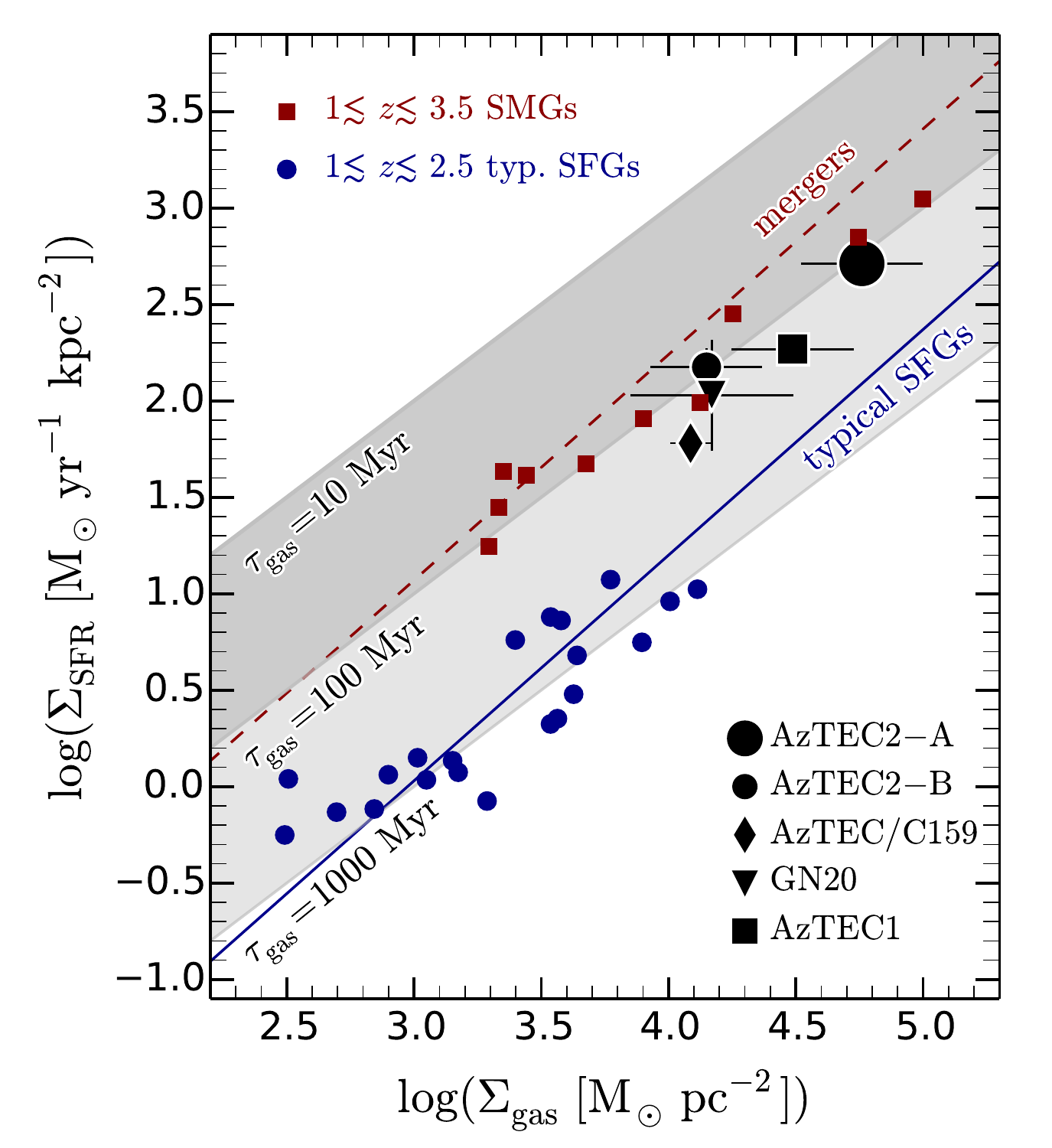}		
		\caption{AzTEC2-A and AzTEC2-B in the Kennicutt-Schmidt  $(\Sigma_{\rm SFR}-\Sigma_{\rm gas})$ plane. For comparison, the massive, star-forming disks  at $z\sim4.5$ GN20 \citep{carilli10, hodge12},  AzTEC/C159 \citep{jones17, jimenezandrade18}, and AzTEC1 \citep{tadaki18} are also shown, along with a compilation of typical SFGs over $1\lesssim z \lesssim 2.5$ and SMGs over $1\lesssim z \lesssim 3.5$ \citep[][and references therein]{genzel10}. In estimating $\Sigma_{\rm SFR}$ and $\Sigma_{\rm gas}$ of $z\sim4.5$ star-forming disks we use their effective radius  from dust continuum emission  and adopt $\alpha_{\rm CO}=2.5\rm \,M_\odot\,K^{-1}\,km^{-1}\,s\,pc^{-2}$, except for AzTEC/C159 for which $\alpha_{\rm CO}=4\rm \,M_\odot\,K^{-1}\,km^{-1}\,s\,pc^{-2}$ has been previously constrained. The horizontal error bars also take into account the  $\Sigma_{\rm gas}$ range given by an $\alpha_{\rm CO}$ varying over $0.8-4.3\rm \,M_\odot\,K^{-1}\,km^{-1}\,s\,pc^{-2}$.    The solid blue (dashed red) line illustrates the KS relation for typical SFGs (mergers) over the redshift range $1\lesssim z \lesssim 3$ derived by \citet{genzel10}. The gray diagonal lines show the $\Sigma_{\rm SFR}$ required to consume the available gas reservoirs within a gas depletion time-scale ($\tau_{\rm gas}$) of 10 Myr (upper), 100 Myr (middle), and 1000 Myr (lower). The gray shaded regions illustrate the $\tau_{\rm gas}$ range for galaxies with (from top-to-bottom) high-to-low SFE. }
		\label{fig:aztec2_ks}
	\end{centering}
\end{figure}

We  use the   spatial extent  of the dust continuum emission from our  ALMA observations { (see \S \ref{subsec:line_detections}) } to infer the galaxy-averaged  SFR surface density: $\Sigma_{\rm SFR}\equiv {\rm SFR}/(2\pi R^2_{\rm eff})$. The effective radius containing half of the total emission, $R_{\rm eff}$, is approximated as $R_{\rm eff}\sim {\rm FWHM}/2.430$ \citep{murphy17}, assuming an exponentially declining surface brightness distribution as observed in disk galaxies. 
In order to approximate the galaxy-averaged molecular gas surface density  ($\Sigma_{\rm gas}\equiv{M_{\rm gas}}/(2\pi R^2_{\rm eff})$),  we use our $M_{\rm gas}$   estimate based on $L^\prime_{\rm CO(1\to0)}$ (\S \ref{sec:gascontent}) and  $\alpha_{\rm CO}=2.5\rm \,M_\odot\,K^{-1}\,km^{-1}\,s\,pc^{-2}$. Although the rotation-dominated gas disk of AzTEC2-A  favors  a higher   $\alpha_{\rm CO}$ value   \citep[e.g., ][]{papadopoulos12b, papadopoulos12, jimenezandrade18}, we require more robust constraints (e.g., dynamical mass, metallicity, excitation conditions) to validate this scenario.  We thus consider the uncertainties of $\Sigma_{\rm gas}$ associated to the unknown $\alpha_{\rm CO}$ value by illustrating in Fig. 6  the $\Sigma_{\rm gas}$ range  given by   $\alpha_{\rm CO}$ varying from 0.8 to $ 4.3\rm \,M_\odot\,K^{-1}\,km^{-1}\,s\,pc^{-2}$. 
  We also recall that in deriving  $L^\prime_{\rm CO(1\to0)}$, and hence $M_{\rm gas}$,   we have assumed typical SMG-like gas excitation conditions, as previously observed in other massive, highly star-forming rotating disks like GN20 and AzTEC/C159  \citep{carilli10, hodge12, jimenezandrade18}.  By combining our  $\Sigma_{\rm gas}$  and  $\Sigma_{\rm SFR}$  estimates  we find that  AzTEC2-A lies at the upper end of the Kennicutt-Schmidt relation for typical SFGs \citep[Fig. \ref{fig:aztec2_ks}][]{genzel10}. This is consistent with the properties of the  massive, star-forming disks GN20 \citep{carilli10, hodge12} and AzTEC/C159 \citep{jones17, jimenezandrade18} at similar redshift.  These systems  arise as scaled (more active) versions of star-forming disks at lower redshifts \citep[$1\lesssim z \lesssim 2.5$ e.g., ][and references therein]{daddi10, daddi10b, genzel10}. Qualitatively, this is in agreement with the systematically higher  gas fractions \citep[e.g., ][]{genzel15, schinnerer16, tacconi18, liu19b}  and enhanced specific SFR of galaxies with increasing redshift \citep[e.g., ][]{karim11, speagle14, schreiber15, lehnert15}.\\

Fig. \ref{fig:aztec2_ks} also suggests that massive, star-forming disks like AzTEC2-A, GN20, and AzTEC/C159 seem to harbor a systematically  lower gas depletion time-scale  than their analogs at lower redshifts; which is compatible with the redshift evolution of $\tau_{\rm gas}$ of massive MS galaxies \citep[e.g., ][]{saintonge13, schinnerer16,  tacconi18, liu19b}. Such an efficient regime of star formation, that approaches that of merger-driven starbursts (Fig. \ref{fig:aztec2_ks}), could be explained by the turbulent ISM and rapid dynamical evolution that characterize high-redshift disks \citep[e.g., ][]{forster-schreiber09, bournaud12, swinbank12}. In this context, the enhanced stellar birth rate of early star-forming disks is a result of their inherent large gas reservoirs and --to some extent-- a higher star formation efficiency ($\equiv 1/\tau_{\rm gas}$). 
Numerical simulations  predict  that at $z>4$ the enhanced gas accretion from the cosmic web can maintain a gravitationally unstable gas-rich disk, which breaks into giant clumps and forms stars at a high rate \citep[][]{bouche07, hodge12, romano-diaz14}. Additionally,   star formation in  AzTEC2-A might be further enhanced due gravitational interaction (torques) with its (minor) companion galaxy, AzTEC2-B, during the ongoing pre-coalescence phase -- as inferred from hydrodynamic merger simulations \citep[e.g., ][]{cox08, moreno15} and observations of galaxy pairs \citep{scudder12}.  Although evidence of such tidal interactions could be inferred from an asymmetric (perturbed) velocity field of the gas \citep[e.g., ][]{kronberger07}, the incomplete coverage of the [C\,II] velocity field of AzTEC2-A prevent us from confirming this scenario.  \\

\section{Summary}

We have used multi-wavelength spectroscopic and photometric  data to constrain the redshift and  conditions for star formation in AzTEC2:  the second brightest SMG (at 1.1 and 2\,mm) in the   COSMOS field. Our results are listed below:

\begin{itemize}
    \item AzTEC2 splits into two components (AzTEC2-A
and AzTEC2-B) for which we detect  $^{12}$CO(5$\to$4) and [CII] line emission, leading to a redshift of $4.626\pm0.001$ and $4.633\pm0.001$ for AzTEC2-A and AzTEC2-B, respectively;

	\item The emission of AzTEC2-A and AzTEC2-B is mildly magnified by a foreground, massive elliptical galaxy at $z=0.34$ located at $\sim4$\,arcsec to the south of AzTEC2-A,B. We estimate a magnification factor of $\mu_{\rm A}=1.5$ and $\mu_{\rm B}=1.35$  for AzTEC2-A and AzTEC2-B, respectively;

	\item Based on the $^{12}$CO(5$\to$4) line emission of AzTEC2-A we have derived $\mu_{\rm A}\,L^{\prime}_{\rm CO(1\to0)}=(12.8\pm2.4)\times10^{10}\,{\rm K\,km\,s}^{-1}\,{\rm pc}^{2}$, implying  a molecular gas mass of $\mu_{\rm A} M_{\rm gas}=(1.0-5.5)\times10^{11}\,M_\odot$.   The   FIR luminosity of AzTEC2-A  leads to $\mu_{\rm A} {\rm SFR}=(2880\pm 140)\,M_\odot\,{\rm yr}^{-1}$,   $L_{\rm IR}/L^{\prime}_{\rm CO(1\to0)}$ ratio of $220\pm50 \,L_\odot({\rm K\,km\,s}^{-1}\,{\rm pc}^{2})^{-1}$, and  $\tau_{\rm gas}=(35-190)\,{\rm Myr}$;  
	
	\item Correspondingly, for AzTEC2-B we have found that $\mu_{\rm B}\,L^{\prime}_{\rm CO(1\to0)}=(3.1\pm0.7)\times10^{10}\, {\rm K\,km\,s}^{-1}\,{\rm pc}^{2}$,  $\mu_{\rm B}\,M_{\rm gas}=(0.25-1.3)\times10^{11}\,M_\odot$,    $\mu_{\rm B}\,{\rm SFR}=960\pm 45\,M_\odot\,{\rm yr}^{-1}$,   $L_{\rm IR}/L^{\prime}_{\rm CO(1\to0)}=310\pm80	\,L_\odot({\rm K\,km\,s}^{-1}\,{\rm pc}^{2})^{-1}$,  and $\tau_{\rm gas}=(25-140)\,{\rm Myr}$;

	\item We have revealed a rotation-dominated [CII] disk in AzTEC2-A, with an intrinsic (de-projected) rotational velocity of  $v_{\rm rot}(i=39^\circ)=660\pm130$\,km\,s$^{-1}$,  velocity dispersion of $\sigma \lesssim100$\,km\,s$^{-1}$ and dynamical mass of $M_{\rm dyn}(i=39^\circ)=2.6^{+1.2}_{-0.9}\times10^{11}\,M_\odot$.  
	
\end{itemize}

Our results indicate that AzTEC2-A hosts a massive, rotation-dominated disk where star formation occurs at intense levels. This indicates that even  disk galaxies, that harbor vast gas reservoirs, could sustain intense star formation activity that resembles that of merger-driven SMGs.  This supports the emerging consensus whereby the population of single-dish selected SMGs is rather heterogeneous, including both interacting  systems and  galaxies that form stars through a smoother mode of star formation sustained by cold gas accretion.
   A more systematic study of high-redshift  star-forming disks is required to verify this scenario, allowing us to probe their properties within  the  framework of the cold and merger mode of star formation  in the early Universe. \\

\acknowledgments
We thank the reviewer for their careful reading of the manuscript and their constructive comments. 
We thank Justin Spilker for  assistance with the gravitational lens modeling. 
E.F.J.A,  B.M.,  E.R.D., A.K., K.C.H., and F.B.  acknowledge  support of the Collaborative Research Center 956, subproject A1 and C4, funded by the Deutsche Forschungsgemeinschaft (DFG). CMC thanks the National Science Foundation for support through grants AST-1714528 and AST-1814034, and acknowledges support from the Research Corporation for Science Advancement from a 2019 Cottrell Scholar Award sponsored by IF/THEN, an initiative of Lyda Hill Philanthropies.  CMC and JAZ also thank the University of Texas at Austin College of Natural Sciences for support. DL acknowledges support and funding from the European Research Council (ERC) under the European Union's Horizon 2020 research and innovation program (grant agreement No. 694343). I.A. acknowledges support from the  CONACyT projects FDC-2016-1828 and CB-2016-281948. A.M. acknowledges support from the CONACyT project A1-S-45680. This paper makes use of the following ALMA data: ADS/JAO.ALMA 2015.1.00568.S. ALMA is a partnership of ESO (representing its member states), NSF (USA) and NINS (Japan), together with NRC (Canada), MOST and ASIAA (Taiwan), and KASI (Republic of Korea), in cooperation with the Republic of Chile. The Joint ALMA Observatory is operated by ESO, AUI/NRAO, and NAOJ. This work is based on observations carried out under project number W18EU with the IRAM NOEMA Interferometer. IRAM is supported by INSU/CNRS (France), MPG (Germany) and IGN (Spain). The National Radio Astronomy Observatory is a facility of the National Science Foundation operated under cooperative agreement by Associated Universities, Inc. 

%

\vspace{5mm}
\facilities{Atacama Large Millimeter Array (ALMA), NOrthern Extended Millimeter Array (NOEMA)}





\appendix

\section{Properties of AzTEC2}
\begin{table*}[h!]
	\begin{center}
		\caption{Properties of AzTEC2}
		{\small
			\begin{tabular}{ l  l c c c c }
				\hline \hline 
				\textbf{Properties} &  \textbf{Units}  &  \multicolumn{2}{|c|}{\textbf {AzTEC2-A}}  &  \multicolumn{2}{|c|}{\textbf {AzTEC2-B}} 
				\\[0.5ex]
				\hline 
				&   & $^{12}$CO(5$\to$4)   &  [CII] & $^{12}$CO(5$\to$4)   & [CII]     \\[0.5ex]
				\hline  \\
				FWHM & km\,s$^{-1}$ & 890 $\pm$ 150   & \dots  &  650 $\pm$ 150  &  \dots   \\[0.5ex]
				Peak flux  & mJy/beam & 1.2 $\pm$ 0.2    & $13\pm1$  & 0.4 $\pm$ 0.1 & $6\pm2$   \\[0.5ex]
			   	$\mu$\,Integrated flux & Jy\,km\,s$^{-1}$ & 1.07 $\pm$ 0.06$^{a}$   &   $10.9\pm1.2$   & 0.26 $\pm$ 0.03$^{a}$  & $4.2\pm0.9$   \\[0.5ex]
				Central frequency &   GHz  &   102.43 $\pm$ 0.03   & $337.8\pm0.2$  & 102.30 $\pm$ 0.02  & $337.4\pm0.2$    \\[0.5ex]
				\hline 
				$z^{b}$ &  $\dots$ &  \multicolumn{2}{c}{ 4.626 $\pm$ 0.001} & \multicolumn{2}{c}{  4.633 $\pm$ 0.001}    \\[0.5ex]
				$\mu$ &  $\dots$ &  \multicolumn{2}{c}{ 1.5} & \multicolumn{2}{c}{1.35}    \\[0.5ex]				
				RA, Dec & hh:mm:ss.sss, dd:mm:ss.ss  &  \multicolumn{2}{c}{10:00:08.042 $+$02:26:12.19} &  \multicolumn{2}{c}{10:00:07.842 $+$02:26:13.32}    \\ [0.5ex]

				$\sqrt{\mu}{\rm FWHM}^{\rm major-axis}_{\rm [CII]}$& arcsec/kpc  &   \multicolumn{2}{c}{$0.70\pm0.12$/$4.6\pm1.0$} &    \multicolumn{2}{c}{$0.53\pm0.18$/$3.4\pm1.2$}  \\[0.5ex]		
				$\sqrt{\mu}\,{\rm FWHM}^{\rm major-axis}_{\rm ^{12}CO(5\to4)}$& arcsec/kpc  &   \multicolumn{2}{c}{$1.2\pm0.4$/$7.8\pm2.6$} &    \multicolumn{2}{c}{$\dots$}   \\[0.5ex]	
				$\sqrt{\mu}\,{\rm FWHM}^{\rm major-axis}_{\rm dust}$& arcsec/kpc  &   \multicolumn{2}{c}{$0.36\pm0.02$/$2.3\pm0.1$} &    \multicolumn{2}{c}{$0.35\pm0.04$/$2.3\pm0.3$}   \\[0.5ex]	
				$\mu\, L_{\rm  IR}$   & $L_{\odot}$  &      \multicolumn{2}{c}{$(2.88\pm0.13)\times 10^{13}$} & \multicolumn{2}{c}{$(9.60\pm0.45)\times  10^{12}$}   \\[0.5ex]
				$\mu\, L_{\rm  FIR}$   & $L_{\odot}$  &      \multicolumn{2}{c}{$(1.20\pm0.06)\times 10^{13}$} & \multicolumn{2}{c}{$(0.40\pm0.02)\times  10^{13}$}   \\[0.5ex]
				$\mu\, L_{\rm  [CII]}$   & $L_{\odot}$  &      \multicolumn{2}{c}{$(3.09\pm0.37)\times 10^{10}$} & \multicolumn{2}{c}{$(1.17\pm0.30)\times  10^{10}$}   \\[0.5ex]

				$\mu\,L'_{\rm CO(5\to4)}$$^{c}$&  K\,km\,s$^{-1}$\,pc$^{2}$ &   \multicolumn{2}{c}{$(4.1\pm0.2)\times10^{10}$} & \multicolumn{2}{c}{$(1.0\pm0.1)\times10^{10}$}    \\[0.5ex]
												
				$\mu\,L'_{\rm CO(1\to0)}$ &  K\,km\,s$^{-1}$\,pc$^{2}$ &   \multicolumn{2}{c}{$(12.8\pm2.4)\times10^{10}$} & \multicolumn{2}{c}{$(3.1\pm0.7)\times10^{10}$}    \\[0.5ex]
				$L_{\rm IR}/L'_{\rm CO(1\to0)}$ & 
				$L_{\odot}$(K\,km\,s$^{-1}$\,pc$^{2}$)$^{-1}$  & \multicolumn{2}{c}{$220\pm50$}  & \multicolumn{2}{c}{$310\pm80$}   \\[0.5ex]
				$\mu\,$SFR   & $M_{\odot}$ yr$^{-1}$  &      \multicolumn{2}{c}{$2880\pm140$} & \multicolumn{2}{c}{$960\pm45$}    \\[0.5ex]
				$\mu\,M_{ \rm gas}$($\alpha_{\rm CO}=0.8, \alpha_{\rm CO}=4.3)$  &  $M_\odot$ &  \multicolumn{2}{c}{$(1.0\pm0.2, 5.5\pm1.0)\times10^{11}$} &  \multicolumn{2}{c}{$(0.25\pm0.05, 1.3\pm0.3)\times10^{11}$}    \\[0.5ex]
			
				$\mu\,M_{ \rm dust}$ &  $M_\odot$ &  \multicolumn{2}{c}{$(3.2\pm0.1)\times10^{9}$} &  \multicolumn{2}{c}{$(1.07\pm0.03)    \times10^{9}$}    \\[0.5ex]

				$\tau_{\rm gas}(\alpha_{\rm CO}=0.8, \alpha_{\rm CO}=4.3)$         &   Myr &  \multicolumn{2}{c}{$(35\pm7, 190\pm33)$} & \multicolumn{2}{c}{$(25\pm6, 140\pm30)$} 	\\[0.9ex]
				\hline
				$\mu\,S_{21{\rm cm}}^{d}$ & mJy  &   \multicolumn{2}{c}{$0.045\pm0.03$} & \multicolumn{2}{c}{$0.039\pm0.02$}   \\[0.5ex]
			
				$\mu\,S_{10{\rm cm}}^{e}$ & mJy  &   \multicolumn{2}{c}{$0.035\pm0.06$} & \multicolumn{2}{c}{$0.025\pm0.06$}   \\[0.5ex]

				$\mu\,S_{2924{\rm \mu m}}$ & mJy  &   \multicolumn{2}{c}{$0.33\pm0.05$} & \multicolumn{2}{c}{$0.09\pm0.02$} \\

				$\mu\,S_{2000{\rm \mu m}}^{f}$ & mJy  &   \multicolumn{4}{c}{$1.09\pm0.22$}  \\

				$\mu\,S_{1100{\rm \mu m}}^{g}$ & mJy  &   \multicolumn{4}{c}{$11.5\pm1.4$}  \\

				$\mu\,S_{887{\rm \mu m}}$& mJy  &   \multicolumn{2}{c}{$13.3\pm0.5$} &    \multicolumn{2}{c}{$4.5\pm0.5$}  \\

				$\mu\,S_{850{\rm \mu m}}^{h}$ & mJy  &   \multicolumn{4}{c}{$15.8\pm1.6$}  \\

				$\mu\,S_{500{\rm \mu m}}^{i}$ & mJy  &   \multicolumn{4}{c}{$37.5\pm3.7$}  \\		
				
				$\mu\,S_{350{\rm \mu m}}^{i}$ & mJy  &   \multicolumn{4}{c}{$30.9\pm3.5$}  \\

				$\mu\,S_{250{\rm \mu m}}^{i}$ & mJy  &   \multicolumn{4}{c}{$24.9\pm2.5$}  \\	
	
				$\mu\,S_{160{\rm \mu m}}^{j}$ & mJy  &   \multicolumn{4}{c}{$17.3\pm7.1$}  \\	
				
				$\mu\,S_{100{\rm \mu m}}^{j}$& mJy  &   \multicolumn{4}{c}{$6.79\pm2.50$}  \\				
		        $\mu\,S_{24{\rm \mu m}}^{k}$& mJy  &   \multicolumn{4}{c}{$0.195\pm0.019$}  
				\\[0.9ex]
				\hline 
				\hline
			\end{tabular}\\
	\tablecomments{$^{a}$These observed flux densities will increase by a factor [1/0.8] when considering the effect of the   CMB.  We use the corrected value, i.e., $\mu\, S^{ {\rm intrinsic}}_{\rm CO(5\to4)}=1.34\pm 10$\,Jy\,km\,s$^{-1}$ and $325\pm 40$\,mJy\,km\,s$^{-1}$ for AzTEC2-A and AzTEC2-B, respectively. $^b$ Redshift derived from $^{12}$CO(5$\to$4) line measurements.  $^{c}$ The effect of the CMB is considered when deriving this value (see \S \ref{sec:gascontent}).  $^{d}$\citet{schinnerer10} $^{e}$\citet{smolcic17}  $^{f}$\citet{magnelli19} $^{g}$\citet{aretxaga11} $^{h}$\citet{geach16b} $^{i}$\citet{oliver12} $^{j}$\citet{lutz11} $^{k}$\citet{lefloch09}.   } }
	\end{center}
	\label{table_aztec2}
\end{table*}

\clearpage 




\bibliography{allrefphdthesis.bib} 



\end{document}